# Ozone Production in Electron Irradiated $CO_2$:$O_2$ Ices

Duncan V. Mifsud,[1,2,†] Zuzana Kaňuchová,[3,†] Sergio Ioppolo,[4,†] Péter Herczku,[2,†] Alejandra Traspas Muiña,[4] Béla Sulik,[2] K. K. Rahul,[2] Sándor T. S. Kovács,[2] Perry A. Hailey,[1] Robert W. McCullough,[5] Nigel J. Mason,[1,†] and Zoltán Juhász[2,†]

1 *Centre for Astrophysics and Planetary Science, School of Physical Sciences, University of Kent, Canterbury CT2 7NH, United Kingdom*

2 *Institute for Nuclear Research (Atomki), Debrecen H-4026, Hungary*

3 *Astronomical Institute, Slovak Academy of Sciences, Tatranska Lomnicá SK-059 60, Slovakia*

4 *School of Electronic Engineering and Computer Science, Queen Mary University of London, London E1 4NS, United Kingdom*

5 *Department of Physics and Astronomy, School of Mathematics and Physics, Queen's University Belfast, Belfast BT7 1NN, United Kingdom*

† Corresponding authors:

| | |
|---|---|
| D. V. Mifsud | dm618@kent.ac.uk |
| Z. Kaňuchová | zkanuch@ta3.sk |
| S. Ioppolo | s.ioppolo@qmul.ac.uk |
| P. Herczku | herczku@atomki.hu |
| N. J. Mason | n.j.mason@kent.ac.uk |
| Z. Juhász | zjuhasz@atomki.hu |

ORCID Identification Numbers

| | |
|---|---|
| D. V. Mifsud | 0000-0002-0379-354X |
| Z. Kaňuchová | 0000-0001-8845-6202 |
| S. Ioppolo | 0000-0002-2271-1781 |
| P. Herczku | 0000-0002-1046-1375 |
| A. Traspas Muiña | 0000-0002-4304-2628 |
| B. Sulik | 0000-0001-8088-5766 |
| K. K. Rahul | 0000-0002-5914-7061 |
| S. T. S. Kovács | 0000-0001-5332-3901 |
| P. A. Hailey | 0000-0002-8121-9674 |
| R. W. McCullough | 0000-0002-4361-8201 |
| N. J. Mason | 0000-0002-4468-8324 |
| Z. Juhász | 0000-0003-3612-0437 |

**Abstract**

The detection of ozone ($O_3$) in the surface ices of Ganymede, Jupiter's largest moon, and of the Saturnian moons Rhea and Dione, has motivated several studies on the route of formation of this species. Previous studies have successfully quantified trends in the production of $O_3$ as a result of the irradiation of pure molecular ices using ultraviolet photons and charged particles (i.e., ions and electrons), such as the abundances of $O_3$ formed after irradiation at different temperatures or using different charged particles. In this study, we extend such results by quantifying the abundance of $O_3$ as a result of the 1 keV electron irradiation of a series of 14 stoichiometrically distinct $CO_2$:$O_2$ astrophysical ice analogues at 20 K. By using mid-infrared spectroscopy as our primary analytical tool, we have also been able to perform a spectral analysis of the asymmetric stretching mode of solid $O_3$ and the variation in its observed shape and profile among the investigated ice mixtures. Our results are important in the context of better understanding the surface composition and chemistry of icy outer Solar System objects, and may thus be of use to future interplanetary space missions such as the ESA *Jupiter Icy Moons Explorer* and the NASA *Europa Clipper* missions, as well as the recently launched NASA *James Webb Space Telescope*.

# 1 Introduction

Ozone ($O_3$) plays an important role in planetary chemistry. On Earth, gaseous $O_3$ is located in the stratosphere and serves as an excellent absorber of short-wavelength ($\lambda$ = 200-315 nm) ultraviolet photons known to cause damage to biomolecules, and thus has important implications for the development and sustenance of life on the planet [1]. In the solid phase, icy $O_3$ has been detected on the surfaces of several outer Solar System moons such as Ganymede in the Jovian system and Rhea and Dione in the Saturnian system [2,3], where it is thought to be an active participant in surface chemistry by virtue of its potent oxidising nature [4]. Conversely, $O_3$ has not been detected in some of the most well-studied comets, such as 1P/Halley and 67P/Churyumov-Gerasimenko, despite relatively high abundances of $O_2$ having been detected in their comae [5,6].

A good understanding of the astrochemical reactions leading to the production of $O_3$ is thus integral to better constraining the chemistry of planetary, lunar, and other Solar System bodies. Accordingly, several laboratory experiments have been performed to explore the formation of $O_3$ as a result of the irradiation of astrophysical molecular ice analogues using ultraviolet photons and charged particles (i.e., ions and electrons). Perhaps the best studied of these ices is $O_2$, whose irradiation by ultraviolet photons, ions, and electrons has been studied extensively and has been shown to produce high yields of $O_3$ [7-15].

Such experiments have demonstrated the dependence of $O_3$ formation in irradiated $O_2$ ices on a number of experimental parameters. For instance, Sivaraman *et al.* [8] demonstrated that lower temperatures are more conducive to the formation of $O_3$ due to higher recombination rates of radiolytically derived oxygen atoms to reform $O_2$ at higher temperatures, thus leaving fewer atoms available to react with $O_2$ and produce $O_3$. Interestingly, there appears to be no dependence of the $O_3$ yield on the dose supplied or the mass of the incident irradiating particle, with irradiations of solid $O_2$ using mono-energetic electrons, protons, and helium, carbon, nitrogen, and oxygen ions all showing that a similar abundance of $O_3$ is produced after a given fluence in each case [7,9,11]. This has been attributed to inelastic stopping interactions being the dominant mechanism of energy transfer in the $O_2$ ice, as all these charged projectiles possess linear energy transfer values which are on the same order of magnitude.

Attention has also been paid to the radiation-induced formation of $O_3$ from molecular ices other than $O_2$, such as $CO_2$ [9,16-21]. Here, $O_3$ is produced as a result of a three-step process which requires that sufficient $O_2$ is first accumulated within the structure of the ice. Temperature has been noted to play a role in this radiolytic chemistry, with increased yields of $O_3$ being recorded on increasing the reaction temperature from 20 to 40 K [21]. On raising the temperature further, however, the total yield of $O_3$ was noted to decline due to increased sublimation-induced losses of the necessary $O_2$ precursor molecules [16,21].

In this paper, we present the results of a systematic study of the 1 keV electron irradiation of a series of 14 stoichiometrically distinct $CO_2$:$O_2$ molecular ices at 20 K, including the two pure end-members. The composition of these ices is particularly relevant to studies of different icy outer Solar System bodies [22]. For example, these species are known to be constituents of the icy nuclei of comets [23,24]. Furthermore, the detection of a tenuous exosphere on the Saturnian moon Rhea composed primarily of $CO_2$ and $O_2$ has led to the suggestion that it is sourced from sputtered or de-gassed surface ices [25]. Such icy outer Solar System bodies are exposed to ionising radiation in the form of the solar wind and giant planetary magnetospheric plasmas. As such, laboratory irradiations of $CO_2$:$O_2$ astrophysical ice analogues are well suited to understanding the chemistry of such celestial bodies.

In this study, we have used mid-infrared spectroscopy to quantify the production efficiency of $O_3$ from the electron irradiated $CO_2$:$O_2$ ices; as well as to determine how the appearance of its mid-infrared asymmetric stretching ($\nu_3$) band varies according to the initial stoichiometric composition of the ice. Understanding such variations is particularly important in light of the fact that $O_3$ is often used as a marker molecule for the presence of $O_2$, which is more difficult to confirm *via* direct spectroscopic observations. Such spectroscopic work could thus greatly aid in the interpretation of data collected by forthcoming interplanetary missions, such as the ESA *Jupiter Icy Moons Explorer* and the NASA *Europa Clipper* missions [26,27]. The recently launched NASA *James Webb Space Telescope* is also anticipated to generate large data-sets of mid-infrared spectroscopic measurements of icy outer Solar System bodies [28], and so could also aid in the detection of $O_3$ in the surface ices of those bodies.

## 2 Experimental Methodology

Experimental work was performed using the Ice Chamber for Astrophysics-Astrochemistry (ICA) located at the Institute for Nuclear Research (Atomki) in Debrecen, Hungary. This set-up has been described in great detail in previous publications [29,30], and so only the most salient features will be presented here. The ICA is an ultra-high vacuum compatible chamber containing a gold-coated sample holder hosting a series of ZnSe deposition substrates which may be cooled to 20 K by a closed-cycle helium cryostat. The temperature of the substrates may be regulated within the 20-300 K range and is measured using two silicon diodes. The pressure in the chamber is typically maintained at a few $10^{-9}$ mbar *via* the combined use of a dry rough vacuum pump and a turbomolecular pump.

The preparation of $CO_2$:$O_2$ astrophysical ice analogues onto the ZnSe substrates was performed *via* background deposition of dosed gases at 20 K. First, $CO_2$ and $O_2$ (both Linde Minican; 99.995%) were introduced into a pre-mixing chamber in the desired stoichiometric ratio, which was determined through standard manometric practices. After being left to equilibrate within the pre-mixing chamber for a few minutes, the gas mixture was dosed into the main chamber *via* an all-metal needle valve at a pressure of a few $10^{-6}$ mbar. Deposition could be followed *in situ* using Fourier-transform mid-infrared transmission absorption spectroscopy (spectral range = 4000-650 $cm^{-1}$; spectral resolution = 1 $cm^{-1}$).

In principle, the thickness $d$ (µm) of a deposited astrophysical ice analogue may be determined spectroscopically by first calculating its molecular column density $N$ (molecules $cm^{-2}$) *via* a modified version of the Beer-Lambert Equation (Eq. 1), and subsequently using this value for $N$ in Eq. 2 [29]:

$$N = \frac{1}{A_\mathrm{v}} \int \tau(\nu)\, \mathrm{d}\nu$$

(Eq. 1)

$$d = \frac{NZ}{\rho N_\mathrm{A}} \times 10^4$$

(Eq. 2)

where $A_\mathrm{v}$ is the integrated band strength constant (cm molecule$^{-1}$) of the mid-infrared absorption band over which Eq. 1 is integrated, $\tau$ ($\nu$) is the optical depth of the ice ($cm^{-1}$), $Z$ is the mass of the molecular constituent whose absorption band was integrated over (g $mol^{-1}$), $\rho$ is the density of the ice (g $cm^{-3}$), and $N_\mathrm{A}$ is the Avogadro constant ($6.02\times10^{23}$ molecule $mol^{-1}$). Eqs. 1 and 2 are valid for pure ices, although their extrapolation to mixed ices is straightforward [29,30]. However, this assumes that all ices in the mixture are infrared active and present absorption bands which may be integrated over.

However, being a homonuclear diatomic molecule, $O_2$ is infrared inactive. Its weak spectral absorption features related to the O–O stretching mode cannot be used to reliably and quantitatively determine the thickness of an ice due to the rather extreme variations in the value of $A_\mathrm{v}$ reported in the literature (for a more complete discussion, see the work of Bennett and Kaiser [7]). Therefore, Eqs. 1 and 2 cannot be used to determine the stoichiometric compositions and thicknesses of our $CO_2$:$O_2$ astrophysical ice analogues. Nevertheless, we are able to provide a reasonable estimate of these parameters for our deposited ices based on the known mixing ratio of the gases in the pre-mixing chamber, as well as quadrupole mass spectrometric measurements of the $CO_2$ and $O_2$ integrated signal curves for the gas mixtures dosed into the main chamber.

We have been able to determine the thickness of the $CO_2$ component of our deposited mixed ices by integrating Eq. 1 over the asymmetric stretching mode ($\nu_3$) at 2344 $cm^{-1}$ and taking $A_\mathrm{v}$ to be $7.6\times10^{-17}$ cm molecule$^{-1}$ [21,31], before using Eq. 2 and taking $\rho$ to be 0.98 g $cm^{-3}$ [32]. By taking $\rho$ to be 1.54 g $cm^{-3}$ for $O_2$ ice [33], and in combination with our spectrometric estimates of the compositions of the deposited ices, we have estimated the total thicknesses of our ices to be in the range 0.20-0.75 µm.

In total, a series of 14 stoichiometrically distinct $CO_2$:$O_2$ ices were deposited onto the ZnSe substrates (Table 1). Once deposited, a mid-infrared absorption spectrum of the ice was collected after which the ice was irradiated using 1 keV electrons at an incidence angle of 36° to the normal. Prior to commencing the experiment, the electron beam current and profile were quantified using the method described previously by Mifsud *et al.* [30]. Our use of a 1 keV electron beam is not only representative of irradiation processes occurring in astrophysical environments such as icy outer Solar System moons, but is also similar to electron beam energies used in previous studies of radiation ice astrochemistry hence allowing for more meaningful comparisons to be made.

The electron beam was scanned over an area of 0.9 $cm^2$, which represents >80% of the area scanned by the mid-infrared spectroscopic beam [29]. Each ice was irradiated for a total of 50 minutes using beam fluxes of $2.0\text{-}2.4\times10^{13}$ $e^-$ $cm^{-2}$ $s^{-1}$, corresponding to total delivered fluences of $6.0\text{-}7.2\times10^{16}$ $e^-$ $cm^{-2}$. The penetration depths of a 1 keV electron in pure $CO_2$ and $O_2$ ices was calculated using the CASINO software [34] and were found to be 65 and 44 nm, respectively (Fig. 1). Given our estimates for the thicknesses of our ices, it may be stated that the incident electrons were effectively implanted into the ice.

**Table 1:** Estimated stoichiometric compositions and thicknesses of the studied ices irradiated by 1 keV electrons.

| Ice | $CO_2$:$O_2$ Ratio | Estimated Molecular Column Density ($10^{17}$ molecules $cm^{-2}$) | | % Content | | Estimated Ice Thickness (μm) |
|---|---|---|---|---|---|---|
| | | $CO_2$ | $O_2$ | $CO_2$ | $O_2$ | |
| 1 | 1:0 | 6.38 | - | 100 | 0 | 0.48 |
| 2 | 73:5 | 7.25 | 0.50 | 94 | 6 | 0.56 |
| 3 | 53:10 | 7.36 | 1.39 | 84 | 16 | 0.60 |
| 4 | 14:5 | 6.90 | 2.47 | 74 | 26 | 0.60 |
| 5 | 2:1 | 7.21 | 3.61 | 67 | 33 | 0.66 |
| 6 | 7:5 | 6.03 | 4.31 | 58 | 42 | 0.60 |
| 7 | 1:1 | 4.10 | 4.09 | 50 | 50 | 0.45 |
| 8 | 5:6 | 3.17 | 3.81 | 45 | 55 | 0.37 |
| 9 | 2:5 | 1.92 | 4.80 | 29 | 71 | 0.31 |
| 10 | 5:14 | 1.23 | 3.46 | 26 | 74 | 0.21 |
| 11 | 1:5 | 1.02 | 5.08 | 17 | 83 | 0.25 |
| 12 | 1:6 | 1.12 | 6.71 | 14 | 86 | 0.31 |
| 13 | 1:50 | 0.41 | 20.69 | 2 | 98 | 0.75 |
| 14 | 0:1 | - | 15.90 | 0 | 100 | 0.55 |

# 3 Results and Discussion

Mid-infrared spectra of the pure $CO_2$ and $O_2$ molecular ices, as well as of the 1:1 binary ice, both before and after 1 keV electron irradiation, are presented in Fig. 2. The onset of irradiation results in the development of new absorption bands due to the formation of new molecules driven by a cascade of tens of thousands of low-energy (<20 eV) secondary electrons [35,36]. The presence of $O_3$ as a radiolytic product was confirmed in each of the investigated ices *via* the detection of its characteristic asymmetric stretching mode ($\nu_3$) at about 1040 $cm^{-1}$. In the mid-infrared region, $O_3$ presents one combination and three fundamental absorption bands [9,37-39]: a bending mode ($\nu_2$) at about 700 $cm^{-1}$, the asymmetric stretching mode ($\nu_3$) at 1040

cm$^{-1}$, a symmetric stretching mode ($\nu_1$) at 1125 cm$^{-1}$, and a combination mode ($\nu_1 + \nu_3$) at 2100 cm$^{-1}$.

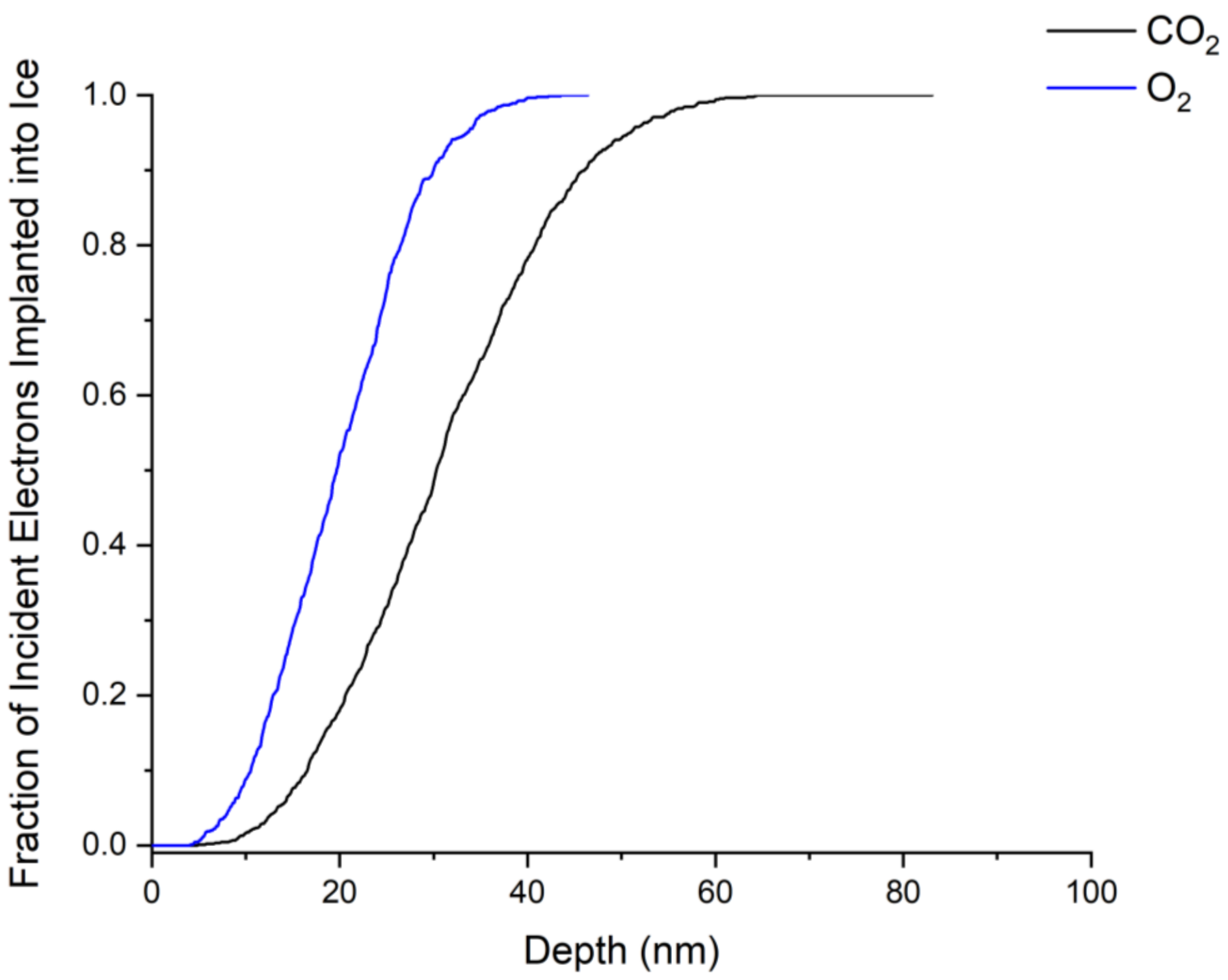

**Fig. 1:** CASINO simulations of the maximum penetration depths of 1 keV electrons in pure $CO_2$ (black trace; 65 nm) and pure $O_2$ (blue trace; 44 nm) ices. Maximum penetration depths for the mixed $CO_2$:$O_2$ ices lie within this range at a value that depends on the stoichiometric composition of the ice. Simulations were performed using an incidence angle of 36° to the normal and using density values of 0.98 and 1.54 g cm$^{-3}$ for $CO_2$ and $O_2$, respectively [32,33].

The radiation chemistry leading to the formation of $O_3$ from irradiated $O_2$ and $CO_2$ ices is straightforward and begins with their electron impact dissociation to yield a supra-thermal oxygen atom (Eqs. 3 and 4) [40,41]. In pure $O_2$ ices, this oxygen atom then combines with a $O_2$ molecule to directly yield $O_3$ (Eq. 5). In the case of pure $CO_2$ ices, the production of $O_3$ is the result of a three-step process which first requires that a sufficient number of oxygen atoms combine to give $O_2$ (Eq. 6), after which the addition of another oxygen atom furnishes the $O_3$ product (Eq. 5). The addition of atomic oxygen to $O_2$ has traditionally been considered to be energetically barrierless [7,8], although Ioppolo *et al.* have proposed a small activation energy barrier of <0.05 eV [42].

$$O_2 \rightarrow 2\ O$$

(Eq. 3)

$$CO_2 \rightarrow CO + O$$

(Eq. 4)

$$O + O_2 \rightarrow O_3$$

(Eq. 5)

$$2\ O \rightarrow O_2$$

(Eq. 6)

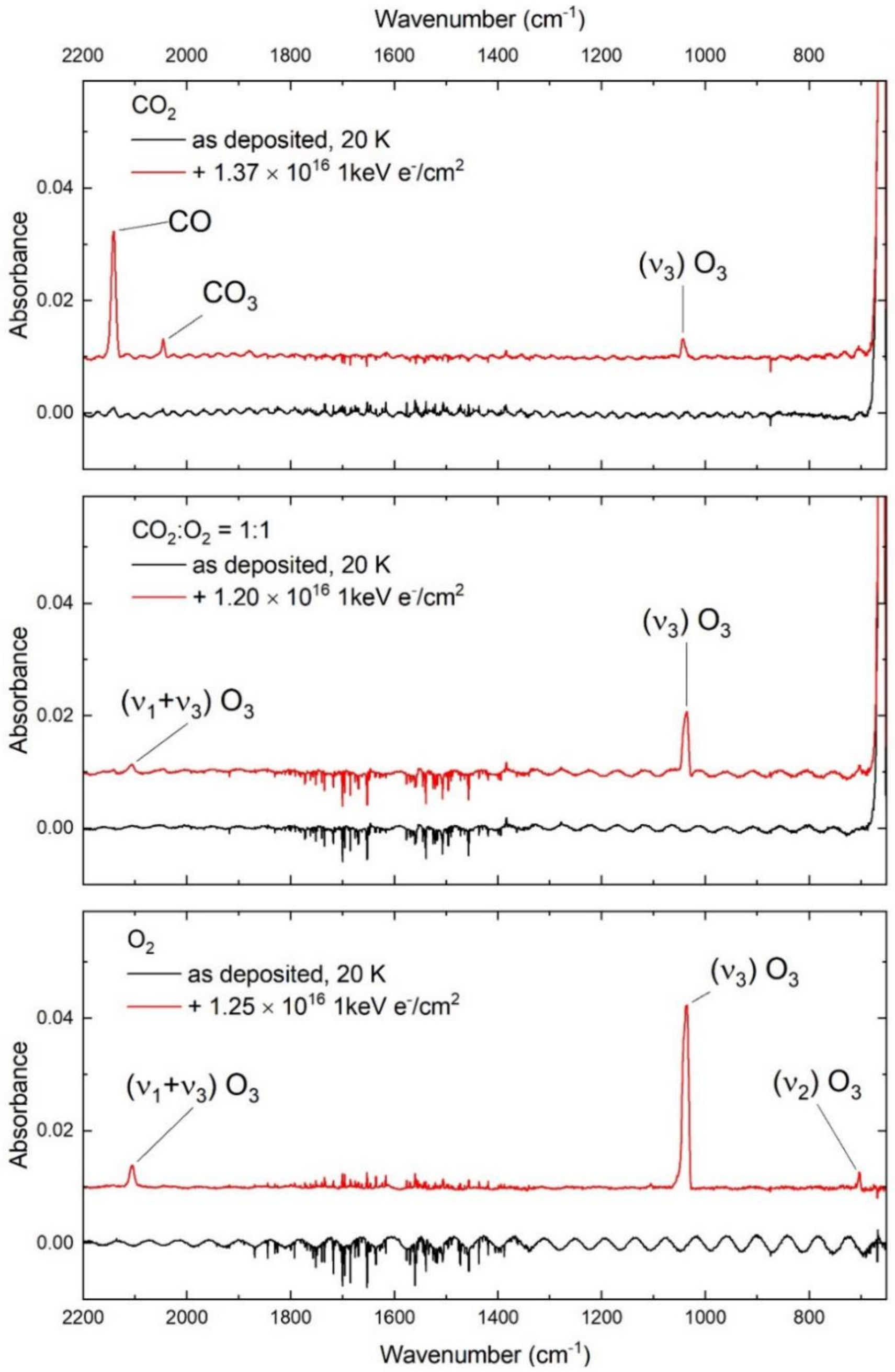

**Fig. 2:** Mid-infrared spectra of pristine and electron irradiated $CO_2$, $O_2$, and 1:1 mixed $CO_2$:$O_2$ astrophysical ice analogues. Products of radiation chemistry are indicated. Spectra are vertically offset for clarity. Oscillations in the baseline of the spectra are the result of interference effects, although these are not anticipated to influence our analyses.

The shape and profile of the $O_3$ asymmetric stretching mode ($\nu_3$) used for its spectroscopic identification offers a chemical insight into its formation. This absorption band is actually a composite structure of three features: the main absorption band due to monomeric $O_3$ centred at 1038 cm$^{-1}$ is sandwiched between two satellite peaks at around 1042 and 1032 cm$^{-1}$ attributed to the ozone dimer [$O_3$…$O_3$] and the ozone-oxygen [$O_3$…O] complexes, respectively [37,38] (Fig. 3). Previous studies have suggested that the positions and profiles of the mid-infrared absorption bands of those species produced *via* the irradiation of mixed molecular ices may depend upon the initial stoichiometric composition of the ice [43-45], and our results suggest that this is also true with regards to the appearance of the asymmetric stretching mode ($\nu_3$) of $O_3$ produced from electron irradiated $CO_2$:$O_2$ ices.

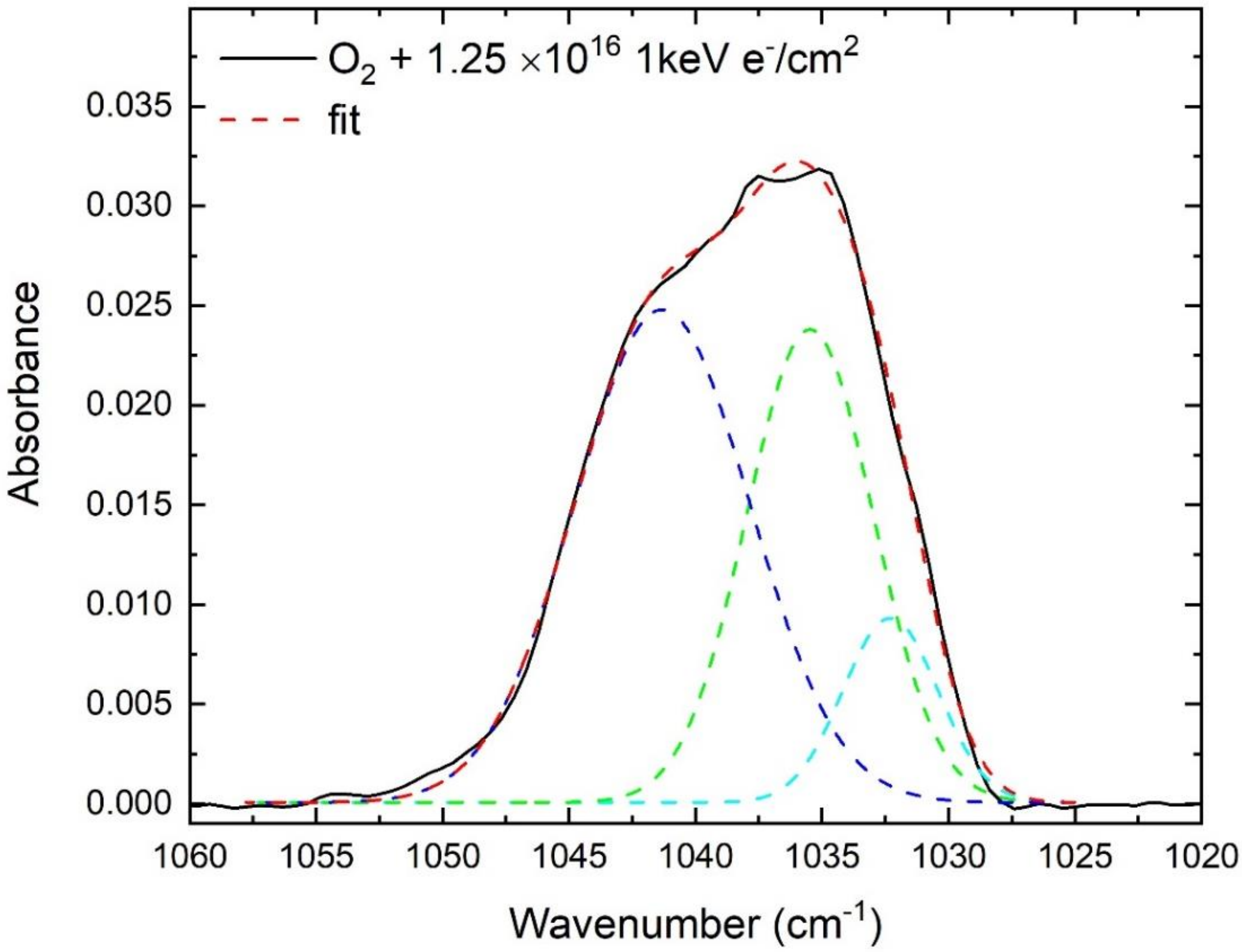

**Fig. 3:** Deconvolution of the $O_3$ asymmetric stretching ($\nu_3$) mode as observed in an electron irradiated pure $O_2$ ice into its constituent Gaussian sub-structures. Similar deconvolutions were performed for all other analysed ices.

Examination of our post-irradiative mid-infrared spectra revealed that the three $CO_2$:$O_2$ mixed ices which were richest in $O_2$ (i.e., the 0:1, 1:50, and 1:6 mixtures) resulted in a $O_3$ asymmetric stretching ($\nu_3$) band which could be deconvoluted into three Gaussian sub-structures, indicating the presence of monomeric $O_3$ and of the [$O_3$…$O_3$] and [$O_3$…O] complexes (Figs. 4 and 5). All irradiated ices with a lower $O_2$ content (or, conversely, a higher $CO_2$ content) exhibited $O_3$ asymmetric stretching ($\nu_3$) bands composed of only two Gaussian sub-structures, with that attributable to the [$O_3$…O] complex being absent in the spectra of these ices. This is in line with the results of previous studies, which have reported the presence of the [$O_3$…O] complex in an electron irradiated pure $O_2$ ice [8] but not in an electron irradiated pure $CO_2$ ice [21].

These observations may be directly related to the molecular environment in each irradiated ice, as free oxygen atoms are far more likely to be in the vicinity of (and thus, be able to complex with) a $O_3$ molecule in an irradiated $O_2$ ice than in an irradiated $CO_2$ ice. This is due to two reasons: firstly, the yield of free supra-thermal oxygen atoms from a dissociated $O_2$ molecule is twice that from a dissociated $CO_2$ molecule leading to a greater abundance of such atoms in a $O_2$ ice. Secondly, the formation of $O_3$ *via* the electron irradiation of $O_2$ ice only suffers from one competing reaction (the reformation of $O_2$) compared to its formation *via* the electron irradiation of $CO_2$ ice, for which many other competing reaction pathways are available which reduce the general abundance of $O_3$ available for co-ordination or complexation [21].

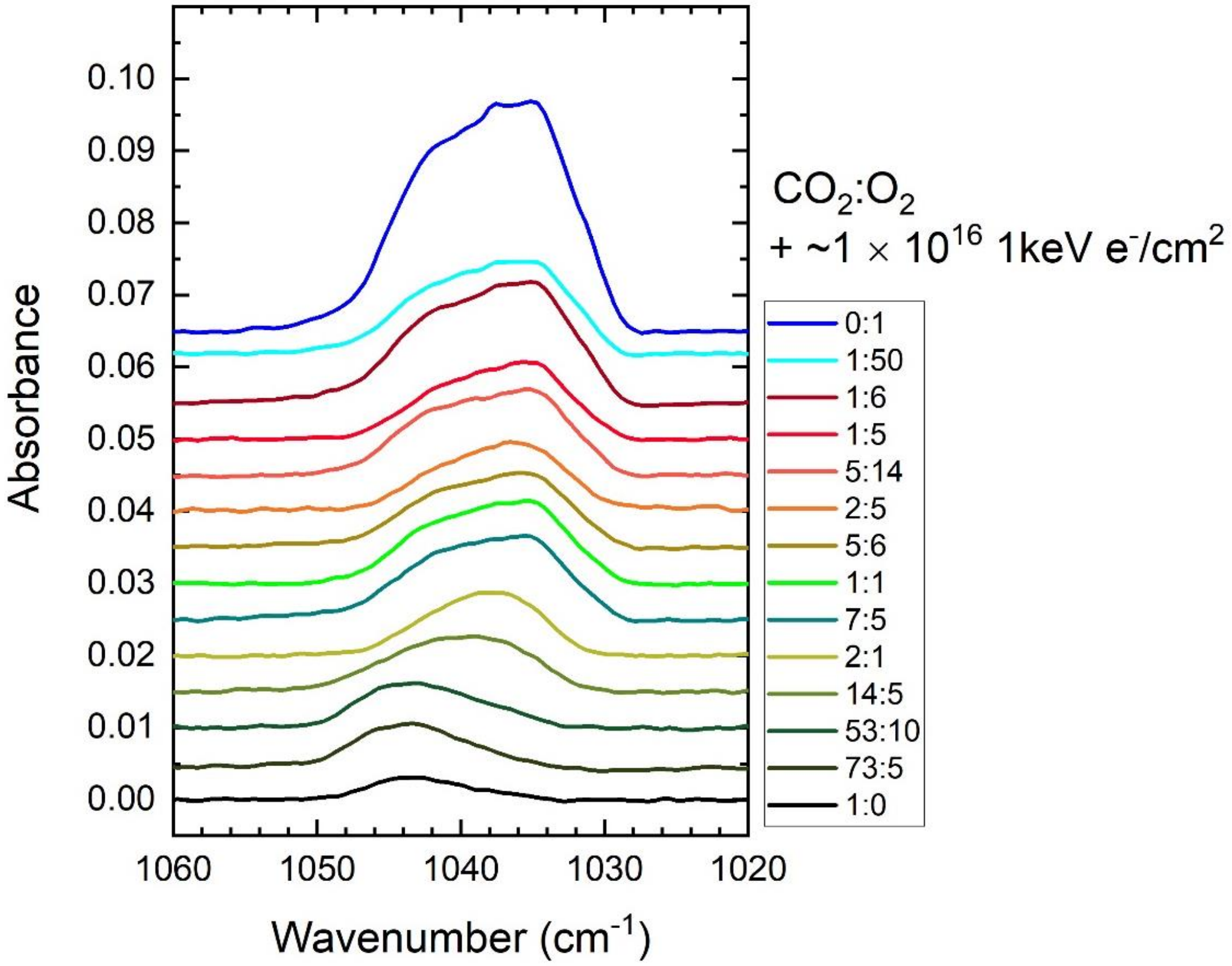

**Fig. 4:** Variation in the shape of the $O_3$ asymmetric stretching ($\nu_3$) mode as observed in 1 keV electron irradiated ($1.3\times10^{16}$ $e^-$ $cm^{-2}$) $CO_2$:$O_2$ mixed ices of different stoichiometric compositions. Note that the scale on the *y*-axis is set to the 1:0 mixed ice, and that all other spectra are vertically offset for clarity.

The positions of the peaks of these band sub-structures were noted to vary from ice to ice, however a pattern is apparent in the case of the monomeric $O_3$ and [$O_3$…$O_3$] peak positions. In the irradiated pure $O_2$ ices, these peaks are respectively located at 1035.5 and 1041.3 $cm^{-1}$, as shown in Fig. 5. On increasing the $CO_2$ content of the ice, these peaks appear to undergo a red-shift to lower wavenumbers, reaching 1034.6 and 1040.5 $cm^{-1}$ in the 1:6 $CO_2$:$O_2$ mixed ice, although we note that this observation may be a consequence of the comparatively higher uncertainties in the peak positions of the two ices richest in $O_2$ (Fig. 5). Further increases in the $CO_2$ content up to a stoichiometric ratio of $CO_2$:$O_2$ = 7:5 do not result in any noticeable changes in the position of these peaks (except for a slight blue-shift of the monomeric $O_3$ peak position in the 2:5 $CO_2$:$O_2$ ice).

On moving to the next $CO_2$:$O_2$ ice in the series (i.e., the 2:1 mixture), however, it is possible to note that although the peak of the Gaussian sub-structure for the [$O_3$…$O_3$] complex does not shift, that for the monomeric $O_3$ sub-structure experiences a blue-shift back to 1035.7 cm$^{-1}$; similar to the position of this peak in the irradiated pure $O_2$ ice. Increasing the $CO_2$ content of the ice even further results in a blue-shift in the position of both peaks, which are located at 1039.9 and 1044.9 cm$^{-1}$ in the 53:10 $CO_2$:$O_2$ ice mixture. Finally, further increasing the $CO_2$ content of the ice results in these peaks undergoing another red-shift, where they are located at 1037.2 and 1043.4 cm$^{-1}$ in the irradiated pure $CO_2$ ice (Fig. 5).

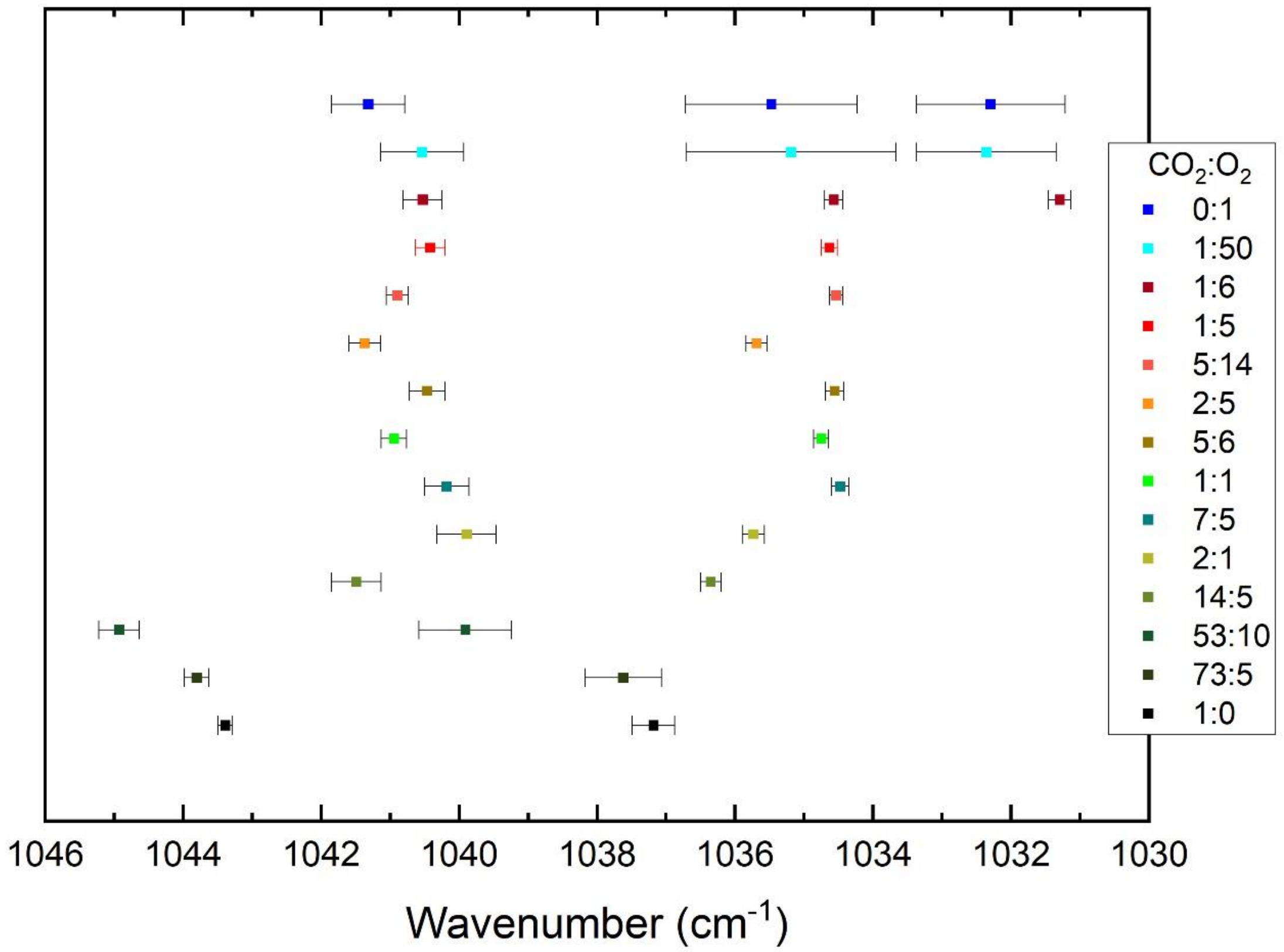

**Fig. 5:** Plot of the peak positions of the Gaussian sub-components of the $O_3$ asymmetric stretching ($\nu_3$) mode in 1 keV electron irradiated $CO_2$:$O_2$ ices of different stoichiometries. Note that a peak for the [$O_3$…O] sub-component was only detected in the three ices with the highest $O_2$ content.

We note that, taken as a whole, the peak position of the $O_3$ asymmetric stretching mode ($\nu_3$) blue-shifts from lower to higher wavenumbers on increasing the $CO_2$ content of the electron irradiated binary $CO_2$:$O_2$ ices, with the exception of the three ices which were richest in $CO_2$ (i.e., the 53:10, 73:5, and the 1:0 mixtures) for which a 'change of direction' in the band peak position shifting is observed (Figs. 4 and 5). It is difficult to provide an exact explanation for these trends, although we note that experimental parameters such optical thickness and molecular environments have been suggested to influence the $O_3$ asymmetric stretching ($\nu_3$) band peak position.

Lastly, we review the $O_3$ productivity of each of the irradiated ices considered in this study. It is evident from Eqs. 3-6 that the successful electron-induced dissociation of a $O_2$ molecule is more efficient at yielding $O_3$ than is the dissociation of a $CO_2$ molecule due to the greater number of reaction pathways available to an irradiated $CO_2$ ice. Indeed, this has been borne out by the results of this study, as the molecular column density of $O_3$ (measured from the peak area of its asymmetric stretching ($\nu_3$) band and taking $A_\nu$ to be 1.4×10$^{-17}$ cm molecule$^{-1}$ [21]) observed in the irradiated pure $O_2$ ice was consistently higher than that observed in the pure $CO_2$ ice throughout irradiation (Fig. 6). This trend was also observed in the electron irradiated mixed $CO_2$:$O_2$ ices.

To better quantify the yield of $O_3$ as a result of the 1 keV electron irradiation of the ices considered in this study, we have defined a percentage formation efficiency based on a series of mass balance calculations which take into account the formation of radiolytic product molecules. Considering first the pure $O_2$ ice: the dissociation of one $O_2$ molecule results in the production of two oxygen atoms which may yield two $O_3$ product molecules as a result of their addition to (or insertion into) two neighbouring $O_2$ molecules. As such, the consumption of three $O_2$ molecules should yield two $O_3$ molecules (Eq. 7). Using a similar logic, the electron-induced dissociation of five $CO_2$ molecules should yield four CO, one $CO_3$, and one $O_3$ molecules (Eq. 8). In the case of the $CO_2$:$O_2$ mixed ices, both mass balance equations are applicable and so Eqs. 7 and 8 may be added together.

$$3\ O_2 \rightarrow 2\ O_3 \qquad \text{(Eq. 7)}$$

$$5\ CO_2 \rightarrow 4\ CO + CO_3 + O_3 \qquad \text{(Eq. 8)}$$

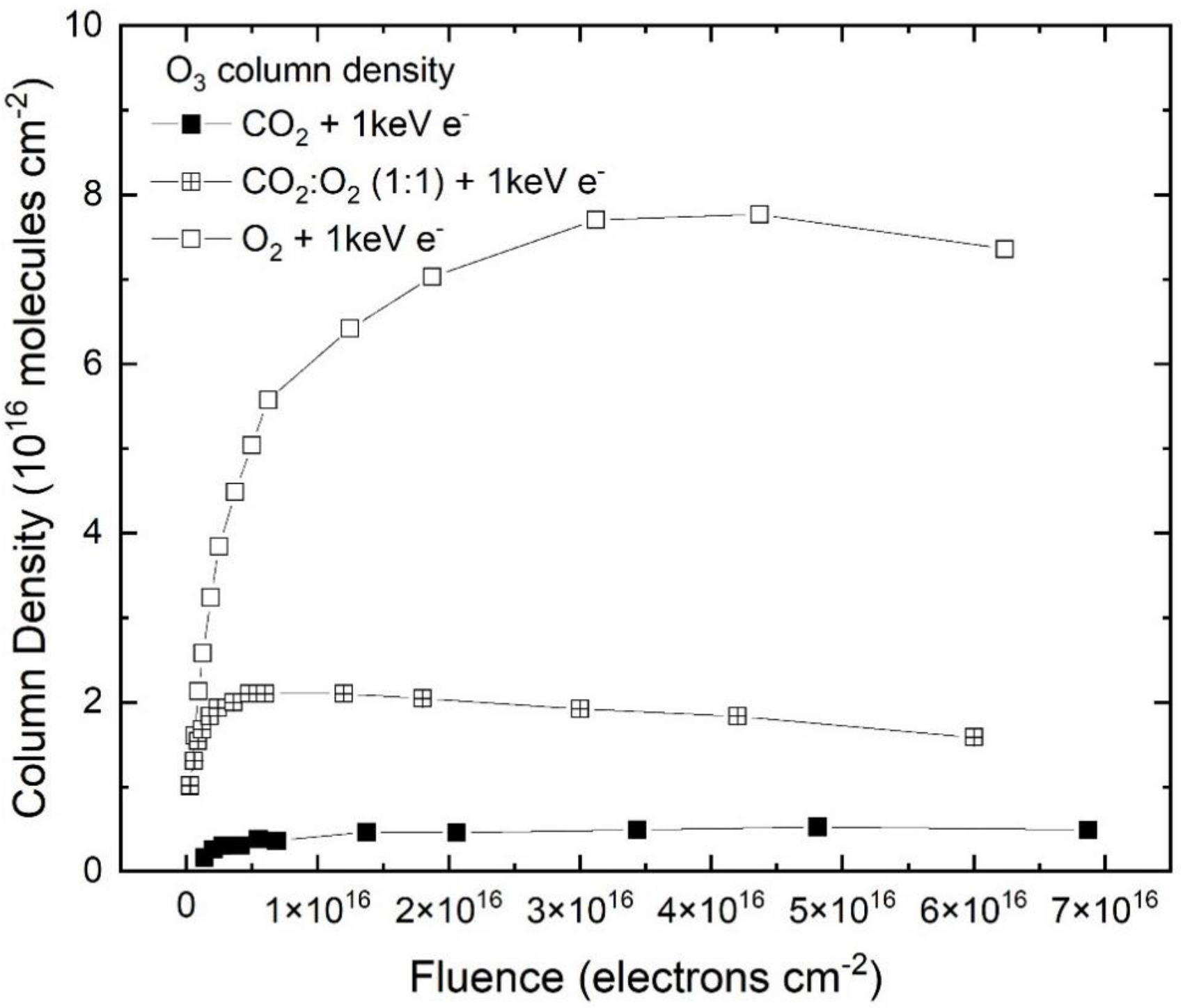

**Fig. 6:** Evolution of $O_3$ column density in electron irradiated pure $O_2$, pure $CO_2$, and 1:1 mixed $CO_2$:$O_2$ ices plotted as a function of electron fluence.

The mass balance relations given in Eqs. 7 and 8, and that used for the mixed ices, are thus indicative of the $O_3$ formation pathways at their most efficient. Hence, by deriving $O_3$ as the subject of the formula in these equations and measuring the initial column densities of $CO_2$ and $O_2$ (in this particular case scaled to the maximum penetration depth of the incident electrons) before irradiation and the column densities of CO and $CO_3$ after a given fluence, the expected $O_3$ column density yield may be computed. By comparing this expected yield to the actual $O_3$ column density measured at that fluence, the percentage formation efficiency of the ice may be calculated fairly simply.

Fig. 7 depicts the $O_3$ formation efficiency of each ice considered in this study at a fluence of $10^{16}$ $e^{-}$ $cm^{-2}$. At this fluence, the column density of $O_3$ is approximately at its maximum in all but three ices: in the pure $O_2$ ice, the $O_3$ column density peaks at a later fluence (Fig. 6) while in the 2:1 and 2:5 mixed $CO_2$:$O_2$ ices, the peak $O_3$ column density occurs at a slightly earlier fluence. Nevertheless, the selected fluence arguably represents the best choice for our analysis of the $O_3$ production efficiency of each irradiated ice. As depicted by the hollow blue circles in Fig. 7, the pure $O_2$ ice was found to have the highest $O_3$ formation efficiency at about 35%. This efficiency then progressively declines as the $CO_2$ content of the ice is raised, falling to about 15% when 25% of the ice is composed of $CO_2$. Interestingly, further increases in the $CO_2$ content of the ice do not noticeably reduce the $O_3$ formation efficiency of the mixed ices as a result of their 1 keV electron irradiation until a $CO_2$ content of about 80% is reached, after which it declines steadily. The pure $CO_2$ ice has a $O_3$ formation efficiency of about 2%.

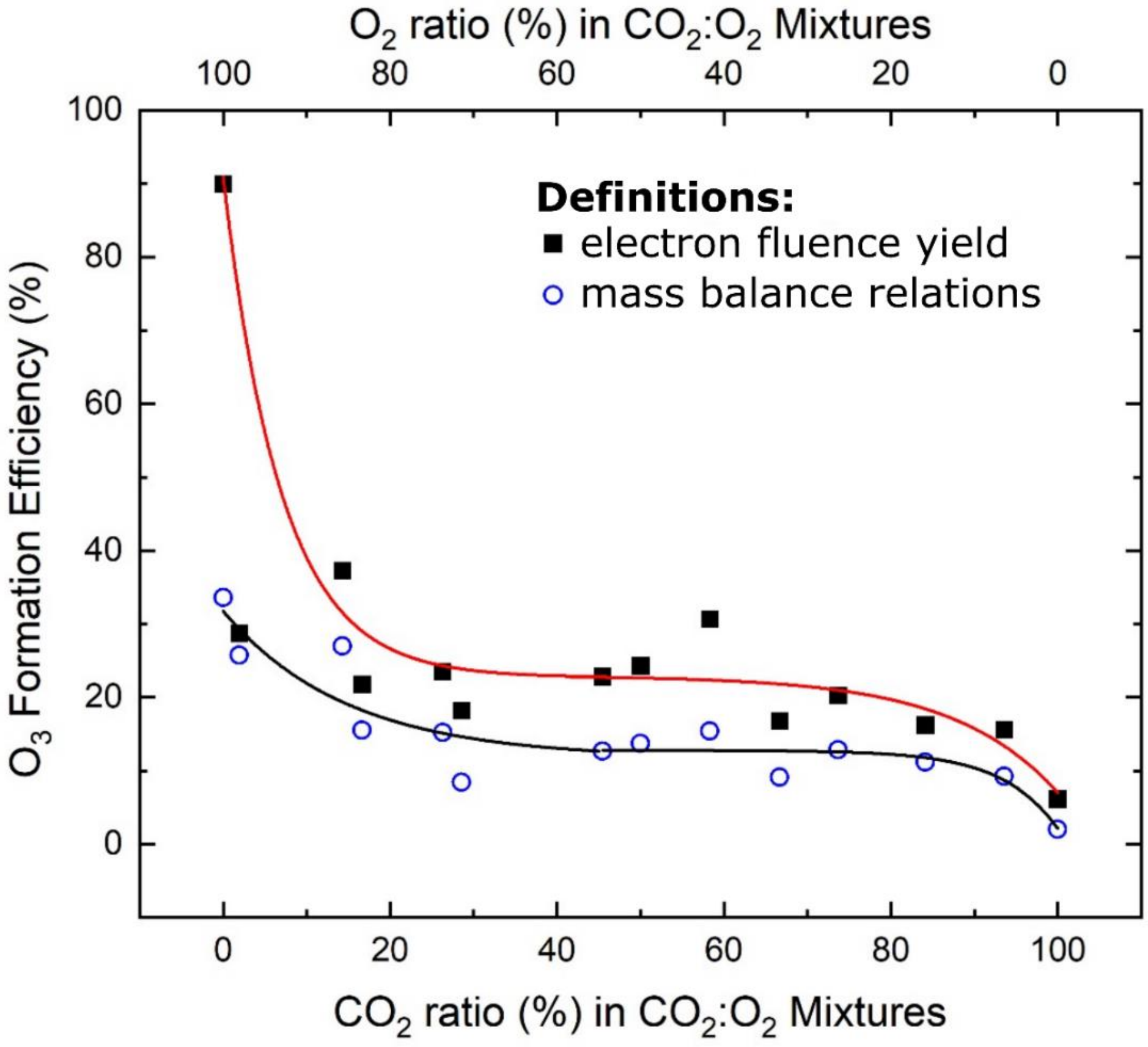

**Fig. 7:** Formation yield of $O_3$ (as a percentage) as a result of the 1 keV electron irradiation of the $CO_2$:$O_2$ ices considered in this study as calculated using the mass balance relations (hollow blue circles) and electron fluence yield (black squares) definitions. Error bars have been omitted, since the uncertainties are anticipated to be significantly lower than the systematic errors arising due to uncertainties associated with the band strength constant $A_v$ used to calculate molecular column densities which, for mixed ices, could be as high as 50%. Each line represents the conjunction of two exponential fits joined at a percentage $CO_2$ content of 45%. Note that, in the case of the electron fluence yield definition, the fit does not include the 1:50 mixture.

In order to ensure that the trend described above is robust and not dependent upon a particular selected definition of $O_3$ formation efficiency, we have re-analysed our data using a different definition for the latter; one based on the electron fluence yield of $O_3$. As a reference, we have considered the case where one primary 1 keV electron dissociates one molecule in the target ice. We acknowledge that such a reference point is not representative of the actual physico-chemical mechanisms by which molecular dissociation in irradiated ices takes place which, as mentioned previously, is the result of the release of tens of thousands of low-energy (<20 eV) electrons [35,36]. However, this reference point is convenient for assessing the $O_3$ formation efficiency of our irradiated ices as a function of primary electron fluence. Using this reference, a maximum of two $O_3$ molecules may be formed per incident electron during irradiation of the pure $O_2$ ice. Conversely, since three electron-induced molecular dissociations are required to generate the necessary number of free oxygen atoms to yield a single $O_3$ molecule during irradiation of a pure $CO_2$ ice, a maximum of one-third $O_3$ molecules per incident electron may be formed. The extension of this definition to the mixed ices is based on their stoichiometric composition.

Hence, using this definition, the reference column density of $O_3$ yielded as a result of electron irradiation is given as the mathematical product of the molecular formation rate per incident electron and the fluence at the maximum $O_3$ abundance in the ice. By comparing this reference column density with the column density of $O_3$ actually measured at this fluence, the percentage formation efficiency of each ice may be computed. As depicted by the black squares in Fig. 7, the $O_3$ formation efficiency trend across ice stoichiometric compositions calculated using the electron fluence yield definition is very similar to that calculated using the mass balance relations definition (depicted by blue hollow circles in Fig. 7). Indeed, most data point pairs are within about 10% of each other, thus indicating the robustness of this trend.

One noticeable discrepancy between the two trends is the $O_3$ formation efficiency of the electron irradiated pure $O_2$ ice, which is calculated to be about 35% when using the mass balance relations and about 90% when using the electron fluence yield definition. It should be noted, however, that the former percentage formation efficiency is very likely underestimated due to it having been calculated using molecular column densities measured when the $O_3$ abundance in the ice had not yet peaked (as explained previously), while the latter most likely indicates a possible upper bound efficiency.

## 4 Implications for Solar System Chemistry

The work presented in this paper is directly applicable to the study of the chemistry of outer Solar System ices, particularly as it relates to the formation of $O_3$. As has been previously noted, $O_3$ has been detected on a number of outer Solar System moons, including several of those of Jupiter and Saturn [2,3]. We note that, on average, the surface temperatures of these moons are higher than the 20 K temperature at which irradiations were performed in this present study. However, temperature gradients across the lunar latitudes are well known, with polar regions being significantly colder than equatorial ones [47]. For example, temperatures as low as 23 K have been reported in the polar regions of Rhea during its long winter [48].

Moreover, several bodies in the Solar System (both inner and outer) with low axial tilts are known to possess permanently shadowed crater regions towards higher latitudes. The temperatures within these regions are significantly lower than the average surface temperature of the planet or moon, and thus allow for the condensation and accumulation of several otherwise volatile molecular species. In the furthest reaches of the Solar System, average surface temperatures are lower still, with volatile molecules such as $N_2$ or $CH_4$ being known to exist as solid ices on the surfaces of Pluto, Charon, and Triton [49,50]. Thus, although our selected irradiation temperature of 20 K is representative of the lowest temperatures in the Solar System, it is still applicable to a wide variety of surface environments and the possible chemistry leading to $O_3$ formation occurring there.

The irradiation conditions presented in this study are also suitable for studying such chemistry, since the 1 keV electron irradiations described simulate the processing that such ices undergo as a result of their interaction with the solar wind or giant planetary magnetospheric plasmas. Indeed, the physico-chemical effects of magnetospheric ion and electron irradiation on the surfaces of the Jovian and Saturnian satellites have been well documented in the literature, and are thought to give rise to the formation of several new and potentially prebiotic molecules [22,51,52]. As such, our experimental conditions allow us to interpret our results in terms of potential magnetospheric plasma-driven $O_3$ synthesis occurring on the surfaces of several of the moons of Jupiter, Saturn, and Uranus.

### 4.1 The Icy Moons of Jupiter

Of the four major (Galilean) moons of Jupiter, the presence of $O_3$ has only been confirmed on Ganymede [2], where the highest molecular abundance is located on the moon's trailing side consistent with observations of icy $O_2$ and with the preferential irradiation by magnetospheric charged particles [53,54]. As such, it is expected that the bulk of the $O_3$ present on Ganymede is sourced from the irradiation of the surface $O_2$ ices.[1] However, $CO_2$ has also been detected at the surface [55], and so it is possible that $O_3$ could also be sourced *via* the irradiation of this molecule [21], although the results of this study combined with the relative distributions of $O_2$ and $CO_2$ at the surface of Ganymede suggest that this contribution is likely to be a minor one.

No $O_3$ has yet been detected on Europa or Callisto, despite the known presence of $O_2$ on both of these moons [56]. The results of this study (as well as those of previous works) demonstrate that the formation of $O_3$ from irradiated $O_2$ ices is rather efficient, and so the non-detection of $O_3$ on these icy moons is somewhat surprising. Loeffler and Hudson [4] have suggested that this lack of $O_3$ may be due to its consumption during the oxidation of other molecular species, such as $SO_2$. Similar arguments have been suggested to explain the non-detection of $O_3$ in cometary and interstellar ices. Forthcoming interplanetary missions to the Jovian moon system, such as the ESA *Jupiter Icy Moons Explorer* and NASA *Europa Clipper* missions [26,27] may detect $O_3$ in observed surface patches of isolated $O_2$ or $CO_2$ exposed to incident magnetospheric plasma, and our spectroscopic results may therefore prove useful in confirming its presence there.

### 4.2 The Icy Moons of Saturn and Uranus

---

[1] It should be noted that, unlike the other Galilean satellites, Ganymede possesses its own magnetosphere, and so the flux of incoming charged particles is expected to be significantly attenuated in equatorial regions. Nevertheless, radiation chemistry is still expected to play a key role in the chemical alteration of the lunar surface in these regions.

All the major moons of Saturn (with the notable exceptions of Titan and Iapetus) are characterised by surfaces dominated by $H_2O$ ices. The detections of $O_3$ on the surfaces of Rhea and Dione [3] therefore provide a challenge, as laboratory studies have thus far been largely unsuccessful in documenting any appreciable yield of $O_3$ as a result of the irradiation of $H_2O$ ices, mainly due to the known catalytic role of OH radicals in the destruction of $O_3$ [57]. However, $CO_2$ is known to be present on the surfaces of both of these moons [58,59], and so its irradiation is a more likely source for the observed $O_3$. This radiation chemistry, driven by the interaction of the lunar surfaces with the Saturnian magnetosphere, is also thought to sustain a tenuous exosphere composed of $CO_2$ and $O_2$ on Rhea [25]. Our irradiations of $CO_2$-rich ices in this present study therefore provide a good analogue of the surface processes occurring on Rhea and Dione.

The synthesis of $O_3$ from electron irradiated $CO_2$-rich $CO_2$:$O_2$ ice mixtures as demonstrated in this study is also applicable to several of the major satellites of Uranus. The surfaces of these moons are composed of a mixture of $H_2O$ ice and dark, carbon-rich refractory material [60]. Laboratory experiments have already demonstrated that the irradiation of such materials gives rise to the formation of CO and $CO_2$ molecules [61,62]. Indeed, $CO_2$ ice has been been firmly detected at the surfaces of Umbriel and Ariel, and tentatively detected at the surfaces of Titania and Oberon [63-65]. $CO_2$ ices there may also be intermixed with smaller quantities of non-native $O_2$ ice sourced from the charged particle and ultraviolet photon irradiation of the surface [66]. As such, our experimental results suggest that there should be some $O_3$ formed at the surfaces of these moons as a result of their interaction with the Uranian magnetosphere.

### 4.3 Ozone on the Moon

Generally speaking, the surface of the Earth's moon is depleted in volatiles [67]. Nevertheless, ices are known to exist there especially within the permanently shadowed regions at the lunar poles. Data collected by the Chandra Altitudinal Composition Explorer (CHACE) instrument on the ISRO *Moon Impact Probe* has shown that oxygen-rich species are prevalent in the lunar exosphere [68], while data from the NASA *Lunar Crater Observation and Sensing Satellite* (LCROSS) has revealed that the same is true for the surface ices [69]. The presence of such species in a radiation environment mediated by the solar wind means that the presence of $O_3$ is also likely (as demonstrated by this and previous studies), although this has yet to be confirmed [70]. As such, our experiments are also representative of possible $O_3$ formation processes in these permanently shadowed regions at the lunar poles.

## 5 Conclusions

In this study, the 1 keV electron irradiation of a series of stoichiometrically distinct $CO_2$:$O_2$ astrophysical ice analogues, including the two pure end-members and 12 binary mixtures, has been studied in detail. Such irradiations are representative of the radiation chemistry occurring in various icy outer Solar System environments. We have been able to successfully quantify the $O_3$ productivity of these ice mixtures as a result of their irradiation, and have determined that the formation efficiency of this species decreases upon the introduction of $CO_2$ to a pure $O_2$ ice. Once the $CO_2$ content of the ice reaches 25%, further additions of $CO_2$ do not noticeably decrease the $O_3$ formation efficiency until a $CO_2$ content of about 70-80% is reached, after which this efficiency declines further.

Using mid-infrared spectroscopy, we have also been able to perform a characterisation of the $O_3$ asymmetric stretching mode ($\nu_3$) in each of the electron irradiated ices. In the three ices richest in $O_2$, this absorption mode may be deconvoluted into three Gaussian sub-structures, indicating the presence of monomeric $O_3$, as well as the [$O_3$…$O_3$] and [$O_3$…O] complexes. In ices containing a higher $CO_2$ content, no spectroscopic evidence for the latter complex was observed. Such results may prove useful in the interpretation of data collected by forthcoming interplanetary missions, and may provide an insight into the formation mechanism of the $O_3$ already observed on several outer Solar System bodies.

## Declaration of Competing Interest

The authors declare that they have no known competing financial interests or personal relationships that could have appeared to influence the work reported in this paper.

## Author Contributions

The experiment was designed by Sergio Ioppolo, Zuzana Kaňuchová, and Alejandra Traspas Muiña and was carried out by Duncan V. Mifsud, Péter Herczku, Béla Sulik, Sándor T. S. Kovács, and Zoltán Juhász. Data analysis was performed by Zuzana Kaňuchová, Duncan V. Mifsud, and Sergio Ioppolo. Duncan V. Mifsud and Zuzana Kaňuchová prepared the manuscript. All authors took part in discussions relating to the interpretation of the results and the improvement of the manuscript.

## Acknowledgements

The authors all gratefully acknowledge funding from the Europlanet 2024 RI which has received funding from the European Union Horizon 2020 Research Innovation Programme under grant agreement No. 871149. The main components of the ICA set-up were purchased with funding from the Royal Society through grants UF130409, RGF/EA/180306, and URF/R/191018. Recent developments of the installation were also supported in part by the Eötvös Loránd Research network *via* grants ELKH IF-2/2019 and ELKH IF-5/2020. Support has also been received from the National Research, Development, and Innovation Fund of Hungary through grant No. K128621.

Duncan V. Mifsud is the grateful recipient of a University of Kent Vice-Chancellor's Research Scholarship. The research of Zuzana Kaňuchová is supported by VEGA – the Slovak Grant Agency for Science (grant No. 2/0059/22) and the Slovak Research and Development Agency (contract No. APVV-19-0072). Sergio Ioppolo acknowledges the Royal Society for financial support. Alejandra Traspas Muiña thanks Queen Mary University of London for doctoral funding.

## References

[1] J. Staehelin, N.R.P. Harris, C. Appenzeller, and J. Eberhard, *Rev. Geophys.*, 2001, **39**, 231. doi: 10.1029/1999RG000059

[2] K.S. Noll, R.E. Johnson, A.L. Lane, D.L. Domingue, and H.A. Weaver, *Science*, 1996, **273**, 341. doi: 10.1126/science.273.5273.341

[3] K.S. Noll, T.L. Roush, D.P. Cruikshank, R.E. Johnson, and Y.J. Pendleton, *Nature*, 1997, **388**, 45. doi: 10.1038/40348

[4] M.J. Loeffler, and R.L. Hudson, *Astrophys. J. Lett.*, 2016, **833**, L9. doi: 10.3847/2041-8213/833/1/L9

[5] A. Bieler, K. Altwegg, H. Balsiger, A. Bar-Nun, J.-J. Berthelier, P. Bochsler, C. Briois, U. Calmonte, M. Combi, J. De Keyser, E.F. van Dishoeck, B. Fiethe, S.A. Fuselier, S. Gasc, T.I. Gombosi, K.C. Hansen, M. Hässig, A. Jäckel, E. Kopp, A. Korth, L. Le Roy, U. Mall, R. Maggiolo, B. Marty, O. Mousis, T. Owen, H. Rème, M. Rubin, T. Sémon, C.-Y. Tzou, J.H. Waite, C. Walsh, and P. Wurz, *Nature*, 2015, **526**, 678. doi: 10.1038/nature15707

[6] M. Rubin, K. Altwegg, E.F. van Dishoeck, and G. Schwehm, *Astrophys. J. Lett.*, 2015, **815**, L11. doi: 10.1088/2041-8205/815/1/L11

[7] C.J. Bennett, and R.I. Kaiser, *Astrophys. J.*, 2005, **635**, 1362. doi: 10.1086/497618

[8] B. Sivaraman, C.S. Jamieson, N.J. Mason, and R.I. Kaiser, *Astrophys. J.*, 2007, **669**, 1414. doi: 10.1086/521216

[9] C.P. Ennis, C.J. Bennett, and R.I. Kaiser, *Phys. Chem. Chem. Phys.*, 2011, **13**, 9469. doi: 10.1039/C1CP20434C

[10] U. Raut, M.J. Loeffler, M. Famá, and R.A. Baragiola, *J. Chem. Phys.*, 2011, **134**, 194501. doi: 10.1063/1.3589201

[11] C.P. Ennis, and R.I. Kaiser, *Astrophys. J.*, 2012, **745**, 103. doi: 10.1088/0004-637X/745/2/103

[12] C.J. Bennett, C.P. Ennis, and R.I. Kaiser, *Astrophys. J.*, 2014, **782**, 63. doi: 10.1088/0004-637X/782/2/63

[13] J. Zhen, and H. Linnartz, *Mon. Not. R. Astron. Soc.*, 2014, **437**, 3190. doi: 10.1093/mnras/stt2106

[14] P. Boduch, R. Brunetto, J.J. Ding, A. Domaracka, Z. Kaňuchová, M.E. Palumbo, H. Rothard, and G. Strazzulla, *Icarus*, 2016, **277**, 424. doi: 10.1016/j.icarus.2016.05.026

[15] S. Ioppolo, Z. Kaňuchová, R.L. James, A. Dawes, N.C. Jones, S.V. Hoffmann, N.J. Mason, and G. Strazzulla, *Astron. Astrophys.*, 2020, **641**, A154. doi: 10.1051/0004-6361/201935477

[16] B. Sivaraman, B.N. Rajasekhar, D. Fulvio, A. Hunniford, R.W. McCullough, M.E. Palumbo, and N.J. Mason, *J. Chem. Phys.*, 2013, **139**, 074706. doi: 10.1063/1.4818166

[17] C.J. Bennett, C.P. Ennis, and R.I. Kaiser, *Astrophys. J.*, 2014, **794**, 57. doi: 10.1088/0004-637X/794/1/57

[18] X.Y. Lv, P. Boduch, J.J. Ding, A. Domaracka, T. Langlinay, M.E. Palumbo, H. Rothard, and G. Strazzulla, *Mon. Not. R. Astron. Soc.*, 2014, **438**, 922. doi: 10.1093/mnras/stt2004.

[19] C. Mejía, M. Bender, D. Severin, C. Trautmann, P. Boduch, V. Bordalo, A. Domaracka, X.Y. Lv, R. Martinez, and H. Rothard, *Nucl. Instrum. Meth. Phys. Res. B*, 2015, **365**, 477. doi: 10.1016/j.nimb.2015.09.039

[20] R. Martín-Doménech, J. Manzano-Santamaría, G.M. Muñoz-Caro, G.A. Cruz-Díaz, Y.-J. Chen, V.J. Herrero, and I. Tanarro, *Astron. Astrophys.*, 2015, **584**, 14. doi: 10.1051/0004-6361/201526003

[21] D.V. Mifsud, Z. Kaňuchová, S. Ioppolo, P. Herczku, A. Traspas Muiña, T.A. Field, P.A. Hailey, Z. Juhász, S.T.S. Kovács, N.J. Mason, R.W. McCullough, S. Pavithraa, K.K. Rahul, B. Paripás, B. Sulik, S.-L. Chou, J.-I. Lo, A. Das, B.-M. Cheng, B.N. Rajasekhar, A. Bhardwaj, and B. Sivaraman, *J. Mol. Spectrosc.*, 2022, **385**, 111599. doi: 10.1016/j.jms.2022.111599

[22] T. Cassidy, P. Coll, F. Raulin, R.W. Carlson, R.E. Johnson, M.J. Loeffler, K.P. Hand, and R.A. Baragiola, *Space Sci. Rev.*, 2010, **153**, 299. doi: 10.1007/s11214-009-9625-3

[23] M. Läuter, T. Kramer, M. Rubin, and K. Altwegg, *Mon. Not. R. Astron. Soc.*, 2019, **483**, 852. doi: 10.1093/mnras/sty3103

[24] M. Combi, Y. Shou, N. Fougere, V. Tenishev, K. Altwegg, M. Rubin, D. Bockelée-Morvan, F. Capaccioni, Y.-C. Cheng, U. Fink, T. Gombosi, K.C. Hansen, Z. Huang, D. Marshall, and G. Toth, *Icarus*, 2020, **335**, 113421. doi: 10.1016/j.icarus.2019.113421

[25] B.D. Teolis, G.H. Jones, P.F. Miles, R.L. Tokar, B.A. Magee, J.H. Waite, E. Roussos, D.T. Young, F.J. Crary, A.J. Coates, R.E. Johnson, W.-L. Tseng, and R.A. Baragiola, *Science*, 2010, **330**, 1813. doi: 10.1126/science.1198366

[26] O. Grasset, M.K. Dougherty, A. Coustenis, E.J. Bunce, C. Erd, D. Titov, M. Blanc, A. Coates, P. Drossart, L.N. Fletcher, H. Hussmann, R. Jaumann, N. Krupp, J.-P. Lebreton, O. Prieto-Ballesteros, P. Tortora, F. Tosi, and T. van Hoolst, *Planet. Space Sci.*, 2013, **78**, 1. doi: 10.1016/j.pss.2012.12.002

[27] C.B. Phillips, and R.T. Pappalardo, *Eos Trans. Am. Geophys. Union*, 2014, **95**, 165. doi: 10.1002/2014EO200002

[28] J.P. Gardner, J.C. Mather, M. Clampin, R. Doyon, M.A. Greenhouse, H.B. Hammel, J.B. Hutchings, P. Jakobsen, S.J. Lilly, K.S. Long, J.I. Lunine, M.J. McCaughrean, M. Mountain, J. Nella, G.H. Rieke, M.J. Rieke, H.-W. Rix, E.P. Smith, G. Sonneborn, M. Stiavelli, H.S. Stockman, R.A. Windhorst, and G.S. Wright, *Space Sci. Rev.*, 2006, **123**, 485. doi: 10.1007/s11214-006-8315-7

[29] P. Herczku, D.V. Mifsud, S. Ioppolo, Z. Juhász, Z. Kaňuchová, S.T.S. Kovács, A. Traspas Muiña, P.A. Hailey, I. Rajta, I. Vajda, N.J. Mason, R.W. McCullough, B. Paripás, and B. Sulik, *Rev. Sci. Instrum.*, 2021, **92**, 084501. doi: 10.1063/5.0050930

[30] D.V. Mifsud, Z. Juhász, P. Herczku, S.T.S. Kovács, S. Ioppolo, Z. Kaňuchová, M. Czentye, P.A. Hailey, A. Traspas Muiña, N.J. Mason, R.W. McCullough, B. Paripás, and B. Sulik, *Eur. Phys. J. D*, 2021, **75**, 182. doi: 10.1140/epjd/s10053-021-00192-7

[31] P.A. Gerakines, W.A. Schutte, J.M. Greenberg, and E.F. van Dishoeck, *Astron. Astrophys.*, 1995, **296**, 810.

[32] R. Luna, M.Á Satorre, M. Domingo, C. Millán, and C. Santonja, *Icarus*, 2012, **221**, 186. doi: 10.1016/j.icarus.2012.07.016

[33] D. Fulvio, B. Sivaraman, G.A. Baratta, M.E. Palumbo, and N.J. Mason, *Spectrochim. Acta A*, 2009, **72**, 1007. doi: 10.1016/j.saa.2008.12.030

[34] D. Drouin, A.R. Couture, D. Joly, X. Tastet, V. Aimez, and R. Gauvin, *Scanning*, 2007, **29**, 92. doi: 10.1002/sca.20000

[35] N.J. Mason, B. Nair, S. Jheeta, and E. Szymańska, *Faraday Discuss.*, 2014, **168**, 235. doi: 10.1039/C4FD00004H

[36] M.C. Boyer, N. Rivas, A.A. Tran, C.A. Verish, and C.R. Arumainayagam, *Surf. Sci.*, 2016, **652**, 26. doi: 10.1016/j.susc.2016.03.012

[37] M. Bahou, L. Schriver-Mazzuoli, and A. Schriver, *J. Chem. Phys.*, 2001, **114**, 4045. doi: 10.1063/1.1342223

[38] H. Chaabouni, L. Schriver-Mazzuoli, and A. Schriver, *J. Phys. Chem. A*, 2000, **104**, 6962. doi: 10.1021/jp0008290

[39] H. Chaabouni, L. Schriver-Mazzuoli, and A. Schriver, *Low Temp. Phys.*, 2000, **26**, 712. doi: 10.1063/1.1312398.

[40] P.C. Cosby, *J. Chem. Phys.*, 1993, **98**, 9560. doi: 10.1063/1.464387

[41] A.S. Morillo-Candas, T. Silva, B.L.M. Klarenaar, M. Grofulović, V. Guerra, and O. Guaitella, *Plasma Sources Sci. Technol.*, 2020, **29**, 01LT01. doi: 10.1088/1361-6595/ab6075

[42] S. Ioppolo, H.M. Cuppen, C. Romanzin, E.F. van Dishoeck, and H. Linnartz, *Astrophys. J.*, 2008, **686**, 1474. doi: 10.1086/591506

[43] M.H. Moore, R.L. Hudson, and R.W. Carlson, *Icarus*, 2007, **189**, 409. doi: 10.1016/j.icarus.2007.01.018

[44] Z. Kaňuchová, P. Boduch, A. Domaracka, M.E. Palumbo, H. Rothard, and G. Strazzulla, *Astron. Astrophys.*, 2017, **604**, A68. doi: 10.1051/0004-6361/201730711

[45] C. Mejía, A.L.F. de Barros, H. Rothard, P. Boduch, and E.F. da Silveira, *Astrophys. J.*, 2020, **894**, 132. doi: 10.3847/1538-4357/ab8935

[46] S.M. Pimblott, and J.A. LaVerne, *Radiat. Phys. Chem.*, 2007, **76**, 1244. doi: 10.1016/j.radphyschem.2007.02.012

[47] Y. Ashkenazy, *Heliyon*, 2019, **5**, e01908. doi: 10.1016/j.heliyon.2019.e01908

[48] C.J.A. Howett, J.R. Spencer, T. Hurford, A. Verbiscer, and M. Segura, *Icarus*, 2016, **272**, 140. doi: 10.1016/j.icarus.2016.02.033

[49] D.P. Cruikshank, T.L. Roush, T.C. Owen, T.R. Geballe, C. De Bergh, B. Schmitt, R.H. Brown, and M.J. Bartholomew, *Science*, 1993, **261**, 742. doi: 10.1126/science.261.5122.742

[50] W.M. Grundy, R.P. Binzel, B.J. Buratti, J.C. Cook, D.P. Cruikshank, C.M. Dalle Ore, A.M. Earle, K. Ennico, C.J.A. Howett, A.W. Lunsford, C.B. Olkin, A.H. Parker, S. Philippe, S. Protopapa, E. Quirico, D.C. Reuter, B. Schmitt, K.N. Singer, A.J. Verbiscer, R.A. Beyer, M.W. Buie, A.F. Cheng, D.E. Jennings, I.R. Linscott, J.W.M. Parker, P.M. Schenk, J.R. Spencer, J.A. Stansberry, S.A. Stern, H.B. Throop, C.C.C. Tsang, H.A. Weaver, G.E. Weigle II, L.A. Young, and the New Horizons Science Team, *Science*, 2016, **351**. doi: 10.1126/science.aad9189

[51] G. Strazzulla, and M.E. Palumbo, *Planet. Space Sci.*, 1998, **46**, 1339. doi: 10.1016/S0032-0633(97)00210-9

[52] L.R. Dartnell, *Astrobiology*, 2011, **11**, 551. doi: 10.1089/ast.2010.0528

[53] J.R. Spencer, W.M. Calvin, and M.J. Person, *J. Geophys. Res. Planet.*, 1995, **100**, 19049. doi: 10.1029/95JE01503

[54] W.M. Calvin, R.E. Johnson, and J.R. Spencer, *Geophys. Res. Lett.*, 1996, **23**, 673. doi: 10.1029/96GL00450

[55] C.A. Hibbitts, R.T. Pappalardo, G.B. Hansen, and T.B. McCord, *J. Geophys. Res. Planet.*, 2003, **108**, 5036. doi: 10.1029/2002JE001956

[56] J.R. Spencer, and W.M. Calvin, *Astron. J.*, 2002, **124**, 3400. doi: 10.1086/344307

[57] P. Boduch, E.F. da Silveira, A. Domaracka, O. Gomis, X.Y. Lv, M.E. Palumbo, S. Pilling, H. Rothard, E. Seperuelo Duarte, and G. Strazzulla, *Adv. Astron.*, 2011, **2011**, 327641. doi: 10.1155/2011/327641

[58] F. Scipioni, F. Tosi, K. Stephan, G. Filacchione, M. Ciarniello, F. Capaccioni, P. Cerroni, and the VIMS Team, *Icarus*, 2013, **226**, 1331. doi: 10.1016/j.icarus.2013.08.008

[59] F. Scipioni, F. Tosi, K. Stephan, G. Filacchione, M. Ciarniello, F. Capaccioni, P. Cerroni, and the VIMS Team, *Icarus*, 2014, **234**, 1. doi: 10.1016/j.icarus.2014.02.010

[60] R.J. Cartwright, J.P. Emery, A.S. Rivkin, D.E. Trilling, and N. Pinilla-Alonso, *Icarus*, 2015, **257**, 428. doi: 10.1016/j.icarus.2015.05.020

[61] V. Mennella, M.E. Palumbo, and G.A. Baratta, *Astrophys. J.*, 2004, **615**, 1073. doi: 10.1086/424685

[62] V. Mennella, G.A. Baratta, M.E. Palumbo, and E.A. Bergin, *Astrophys. J.*, 2006, **643**, 923. doi: 10.1086/502965

[63] W.M., Grundy, L.A. Young, J.R. Spencer, R.E. Johnson, E.F. Young, and M.W. Buie, *Icarus*, 2006, **184**, 543. doi: 10.1016/j.icarus.2006.04.016

[64] W.M. Grundy, L.A. Young, and E.F. Young, *Icarus*, 2003, **162**, 222. doi: 10.1016/S0019-1035(02)00075-1

[65] M.M. Sori, J. Bapst, A.M. Bramson, S. Byrne, and M.E. Landis, *Icarus*, 2017, **290**, 1. doi: 10.1016/j.icarus.2017.02.029

[66] C. Ahrens, H. Meraviglia, and C. Bennett, *Geosci.*, 2022, **12**, 51. doi: 10.3390/geosciences12020051

[67] P.G. Lucey, N. Petro, D.M. Hurley, W.M. Farrell, P. Prem, E.S. Costello, M.L. Cable, M.K. Barker, M. Benna, M.D. Dyar, E.A. Fisher, R.O. Green, P.O. Hayne, K. Hibbitts, C. Honniball, S. Li, E. Malaret, K. Mandt, E. Mazarico, M. McCanta, C. Pieters, X. Sun, D. Thompson, and T. Orlando, *Geochem.*, 2021, doi: 10.1016/j.chemer.2021.125858

[68] R. Sridharan, S.M. Ahmed, T.P. Das, P. Sreelatha, P. Pradeepkumar, N. Naik, and G. Supriya, *Planet. Space Sci.*, 2010, **58**, 947. doi: 10.1016/j.pss.2010.02.013

[69] A. Colaprete, P. Schultz, J. Heldmann, D. Wooden, M. Shirley, K. Ennico, B. Hermalyn, W. Marshall, A. Ricco, R.C. Elphic, D. Goldstein, D. Summy, G.D. Bart, E. Asphaug, D. Korycansky, D. Landis, and L. Sollitt, *Science*, 2010, **330**, 463. doi: 10.1126/science.1186986

[70] B. Sivaraman, B.G. Nair, B.N. Rajasekhar, J.-I. Lo, R. Sridharan, B.-M. Cheng, and N.J. Mason, *Chem. Phys. Lett.*, 2014, **603**, 33. doi: 10.1016/j.cplett.2014.04.021